paper.tex000644 000747 000017 00000021555 06316231166 013431 0ustar00zavlinusers000000 000000 
\documentstyle[epsf,sprocl]{article}

\def\be{\begin{equation}}
\def\ee{\end{equation}}
\def\bea{\begin{eqnarray}}
\def\eea{\end{eqnarray}}

\begin{document}
\title{
CONSTRAINTS ON THE MASS AND RADIUS OF PULSARS
FROM X-RAY OBSERVATIONS OF THEIR POLAR CAPS}
\author{G.G. PAVLOV}
\address{Pennsylvania State University, 525 Davey Lab,\\
University Park, PA 16802, USA}
\author{V.E. ZAVLIN}
\address{Max--Planck--Institut f\"ur Extraterrestrische Physik,\\
D-85740 Garching, Germany}
\maketitle\abstracts{
The properties of X-ray radiation from the polar caps 
predicted by the radio pulsar models depend on the surface
chemical composition, magnetic field and star's mass and radius
as well as on the cap temperature, size and position.
Fitting the radiation spectra and light curves 
with the neutron star atmosphere models enables one 
to infer these parameters.  We present here results obtained 
from the analysis of the soft X-ray radiation of PSR J0437-4715.
In particular, with the aid of radio polarization
data, we put constraints on the pulsar mass-to-radius ratio.
}
\section{Introduction}
Current models of radio pulsars~\cite{cr80,a81,bgi93}
predict a typical polar cap (PC) size 
$R_{\rm pc}\sim (2\pi R^3/Pc)^{1/2}$ and PC temperatures
$T_{\rm pc}\sim 3\times 10^5 - 6\times 10^6$~K,  depending
on the model adopted.  The spectra of the thermal PC radiation
are mainly determined by the
temperature, gravitational acceleration,
magnetic field and
chemical composition of emitting layers (atmospheres). 
The light curves of the pulsed PC radiation depend not only on 
the orientation of the magnetic and rotation axes, but  also on the 
magnetic field and chemical composition which affect
the angular distribution of radiation,
and the neutron star (NS) mass-to-radius ratio, $M/R$, 
which determines the gravitational bending of the photon trajectories. 
The best candidates for the investigation of the PC radiation are nearby, old
pulsars for which both the nonthermal radiation from relativistic
particles and thermal radiation from the entire NS
surface are expected to be negligibly faint.
Here we report results obtained for PSR J0437--4715 ($P=5.75$~ms,
$\tau=P/2\dot{P} = 5\times 10^9$~yr, $B\sim 3\times 10^8$~G, $d\approx 180$~pc).
$ROSAT$ observations of this object
have revealed~\cite{bt93}
smooth pulsations with the pulsed fraction $\sim 25-50\%$
growing with the photon energy.    
Such behavior has been predicted~\cite{zps96}
for the radiation emergent from the NS atmospheres
with low magnetic fields $B < 10^{10}$~G. 
Making use of the low-field NS atmosphere models~\cite{zps96}, 
we model PC radiation with
allowance for the gravitational effects~\cite{zsp95}
and compare the results with observational data.

\section{Results}
Our analysis has shown~\cite{zpbt97}
that both the spectra and the light curves
obtained with  $ROSAT$ and $EUVE$ can be interpreted 
as thermal radiation from two PCs with radii
$R_{\rm pc}=0.8-0.9$~km (comparable to 1.9 km expected from the
simple estimate) covered with hydrogen or helium at a temperature  
$T_{\rm pc}=(8-9)\times 10^5$~K. 
Neither the blackbody nor iron atmosphere models give acceptable
spectral fits.  Those results were obtained for fixed 
$M=1.4 M_\odot$ and $R=10$~km,  
and the angles between the pulsar rotation axis and
the line of sight, $\zeta=40^\circ$, and between 
the rotation and magnetic axes, $\alpha=35^\circ$,
evaluated from the phase dependence of the radio polarization
position angle~\cite{mj95}.  We checked that 
these parameters
do not affect significantly the results of the spectral fits (PC
temperature and size). However, the values of $\alpha$, $\zeta$ and $M/R$
drastically affect the shape of the light curve and the pulsed fraction.

\begin{figure}
\epsfxsize=7.5cm
\epsffile[0 70 400 400]{fig1.ps}
\caption[ ]{
NS mass-radius diagram with the lines of constant values of
$M/R$ (the numbers 
in units of $M_\odot/10~{\rm km}$)
and the $M(R)$ curves for soft ($\pi$) and hard (TI and MF)
equations of state of superdense matter~\cite{st83}.
The shaded region indicates the mass-radius
domain compatible with $\zeta=40^\circ$ and
$\alpha=35^\circ$.
}
\label{fig1}
\end{figure}

To constrain these parameters,
we (i) fitted the $ROSAT$ PSPC count rate spectrum with the
hydrogen atmosphere models on
a grid of $\zeta$, $\alpha$ and $M/R$ values and 
obtained the corresponding set of $R_{\rm pc}$ and $T_{\rm pc}$;
(ii) used this set to
compute the model $ROSAT$ PSPC light curves and 
evaluated their deviations ($\chi^2$ values)
from the observed light curve;
(iii) calculated the confidence regions in the $\zeta$-$\alpha$ plane
at trial values of $M/R$. 
If there were no observational information about the $\alpha$, $\zeta$ values,
the only constraint on the $M/R$ ratio would be 
$M < 1.6 M_\odot (R/10~{\rm km})$, or $R>8.8 (M/1.4 M_\odot)$~km,
at the 99\% confidence level.  If, however, we adopt 
$\zeta=40^\circ$ and $\alpha=35^\circ$, the $M/R$ ratio lies in the range 
$1.4 < (M/M_\odot)/(R/10~{\rm km}) <1.6$ (Fig.~1). 
This means that if the pulsar mass is $M=1.4 M_\odot$, its radius 
is in the range $R=8.8-10.0$~km. 
An alternative set of angles~\cite{gk96}, 
$\zeta=24^\circ$ and $\alpha=20^\circ$, 
yields $(M/M_\odot)/(R/10~{\rm km}) < 0.3$, leading to
very low masses, $M < 0.5 M_\odot$, at any $R$ allowed by the equations
of state.

Thus, the analysis of the PC X-ray radiation
in terms of the NS atmosphere models provides a new tool to 
constrain the NS mass and radius
and the equation state of the superdense
matter.  A similar analysis of X-ray radiation from 
pulsars with strong magnetic fields, $B\sim 10^{12}$~G (e.~g., PSR B1929+10)
would enable one to additionally constrain the magnetic field strength.

\section*{Acknowledgments} 
The work was partially supported through the
NASA grant NAG5-2807, INTAS grant 94-3834 and
DFG-RBRF grant 96-02-00177G. VEZ acknowledges the Max-Planck fellowship.

\end{document}